\documentclass{PoS} 
\usepackage{graphicx}
\usepackage{dcolumn}
\usepackage{bm}
\input{symblibrary.tex}
\usepackage{multirow}
\title{ New Results on Bottom Baryons with the CDF~II Detector 
      }
\ShortTitle{Bottom Baryon Resonances with CDF~II}
\author{Constantino~Calancha\\
        (on behalf of the CDF~Collaboration)\\
        CIEMAT,~Madrid,~Spain\\
        E-mail: \email{calancha@fnal.gov}}
\author{\speaker{Igor~V.~Gorelov}\\
        (on behalf of the CDF~Collaboration)\\
        University~of~New~Mexico,~Albuquerque,~USA\\
        E-mail: \email{gorelov@fnal.gov}}
\abstract{ 
  We present measurements of the masses and widths of four bottom baryon
  resonances, \( \Sgbstpm \), reconstructed in the \( \Lb\pipm \) hadron
  decay modes. The isospin mass splittings for the \( \Sigb \) and
  \( \Sigbst \) multiplets are extracted as well. The analysis is based on
  a data sample corresponding to an integrated luminosity of
  \(\IntL\approx\luminosity \invfb \).
}
\FullConference{35th International Conference of High Energy Physics - ICHEP~2010,\\
		July~22~-~28,~2010,\\
		Paris,~France}
\begin{document}
\begin{figure}[ht]
\begin{minipage}[tb]{0.5\linewidth}
\vspace{-0.12in}
\begin{flushleft}
\includegraphics[width=0.88\textwidth]{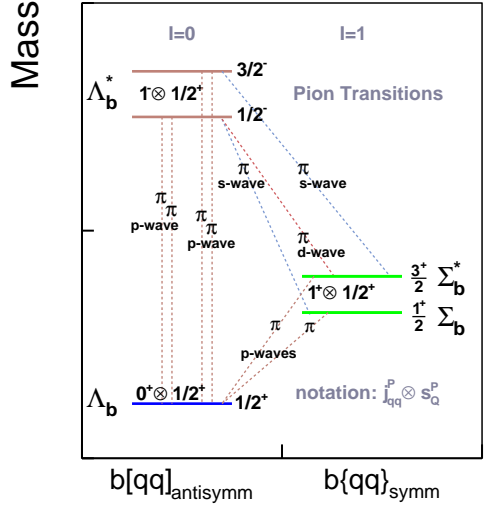} 
\end{flushleft}
\end{minipage}
\begin{minipage}[b]{0.5\linewidth}
\caption{
          The pion transitions of \(\Sgbst\) states and
          a first orbital P-wave excitations \(\Lbst\).  \(\Sgbst\) states
          are the lowest-lying \(S\)- wave states which can decay to the
          \( J^{P} = {\frac{1}{2}}^{+} \) singlet \( \Lb \) via strong
          processes involving soft pion emissions provided sufficient
          phase space is available for a given mode.
  \label{fig:lbmodes}
}
\end{minipage}
\end{figure}

%
%
%
%
%
  The heavy baryons with a single heavy quark are the helium atoms of
  QCD with nucleus as a heavy quark \( Q \) and two orbiting electrons
  as a light di-quark \( q_{1}q_{2} \). The heavy quark in the baryon may
  be used as a probe of a confinement which at least will allow us to
  study a non-perturbative QCD somewhat deeper than we do it with light
  baryons. 
%
  In the experimental analysis presented below we investigate bottom baryon states
  \Sigb,~\Sigbst with the quark content \(\b\{q_{1}q_{2}\} \), whereof a heavy
  quark spin \( S_{b}={\frac{1}{2}}^{+} \) and a spin of the flavor
  symmetric di-quark \(S_{\{q_{1}q_{2}\}}=1^{+} \) constitute two
  isospin \( I=1 \) triplets with the total spin \( J^{P} =
  {\frac{1}{2}}^{+} \) and \( J^{P} = {\frac{3}{2}}^{+} \), see
  Fig.~\ref{fig:lbmodes}.
\par 
  This study follows a discovery of the \Sgbst states made by CDF
  Collaboration in 2006~\cite{:2007rw}.  We intend to confirm the first
  observation of those states using a larger data-sample.  Fitting the
  mass difference spectra of \Sgbstp and \Sgbstm independently and with
  floating background model parameters, we aim to measure masses,
  intrinsic natural widths and isospin mass splittings within \( \Sigb \) 
  and \( \Sigbst \) iso-triplets.  With our measurements we will
  provide the next input to the theoretical community initiating the new
  round of heavy baryon calculations including somewhat newer tasks like
  e.g. natural width estimates.
\par
  Some recent HQET calculations for bottom baryons are done
  in~\cite{hqet}.  
%
  In potential quark model the mass differences like 
  \( {\Sigma}_{Q} - {\Lambda}_{Q} \),
  \( {\Sigma}^{*}_{Q} - {\Sigma}_{Q} \) are accounted largely by hyperfine 
  splittings, hence the mass differences scale as \( 1/{{m}_{Q}} \).
  Some of recent predictions based on potential quark models can be found
  in~\cite{potmodels}. 
%
%
%
  Few of the theoretical predictions on natural widths have been
  published~\cite{widths}.
%
%
%
%
\par
  The analysis is based on an integrated luminosity of \(6.0~\invfb\)
  collected with the \cdf2 detector between March 2002 and February
  2010.  The component of the CDF~II detector~\cite{Acosta:2004yw} most
  relevant for this analysis is the tracking system comprising a
  microvertex silicon detector and a large drift
  chamber~\cite{tracking}. The trigger~\cite{trigger} on displaced
  tracks with \(\pt>2~\gevc\) selects a data sample enriched with
  hadron modes of \b-quarks produced in the detector.
\par
\begin{figure}[ht]
\begin{minipage}[tb]{0.5\linewidth}
\vspace{-0.75in}
\begin{flushleft}
\includegraphics[width=1.0\textwidth]{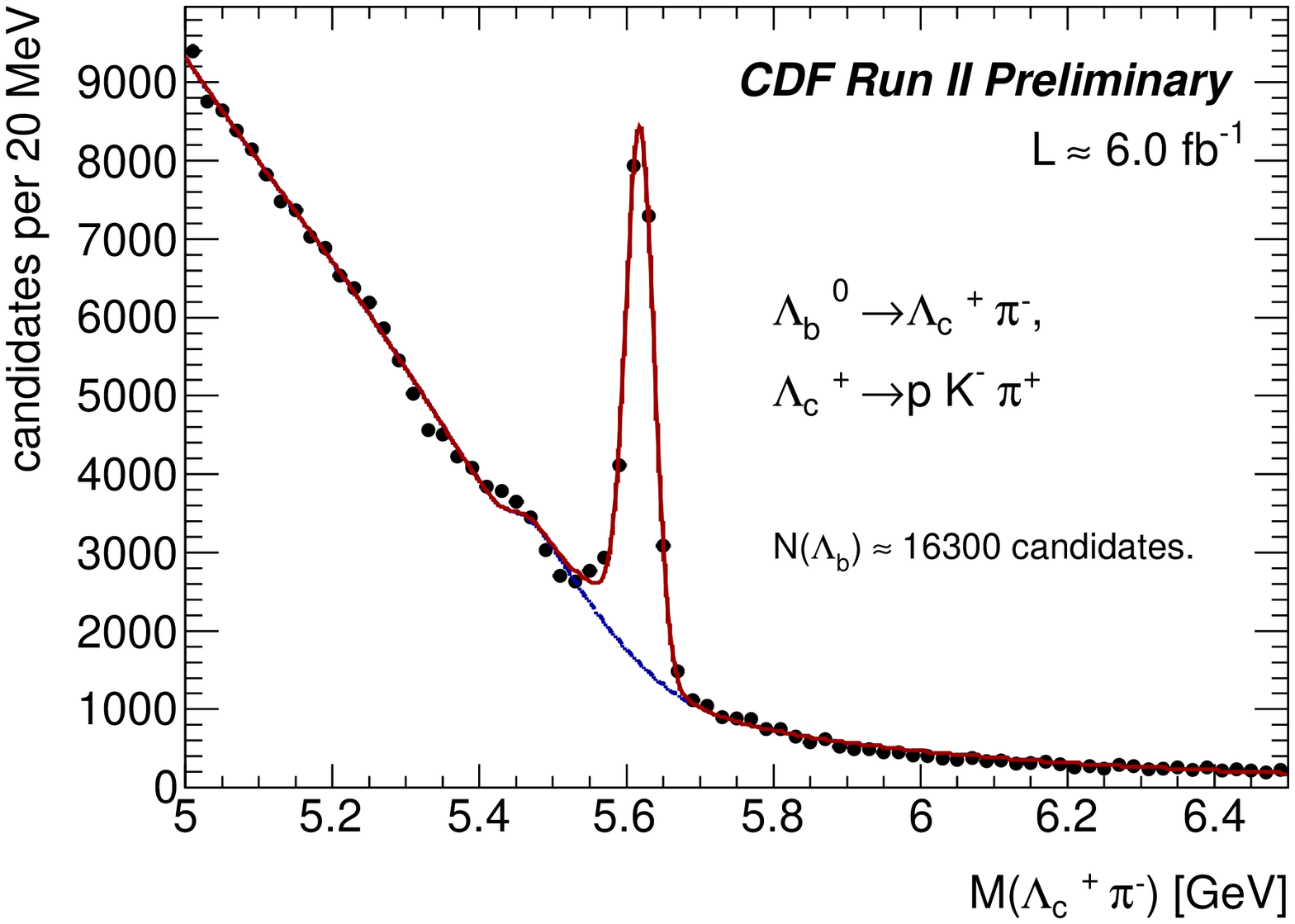} 
\end{flushleft}
\end{minipage}
\begin{minipage}[b]{0.5\linewidth}
\vspace{-0.09in}
\caption{
         \Lb signal reconstructed using the total statistics of 
         \(\IntL = 6\invfb \). In the analysis at least two tracks within
         \( \proton,\,\Km,\,\pip,\,\pim_{b} \) are demanded to meet the
         criteria firing the displaced track trigger. The fitted 
         \( \Lb\to\Lc\pi_{soft} \)- vertex is required to be by at least
         \( 12\cdot\sigma_{\ct} \) away from the primary one, where 
         \(\ct \) is a product of a proper decay time and the speed of light, 
         \( \sigma_{\ct} \) is its vertex fit uncertainty.
  \label{fig:lb_signal}
}
\end{minipage}
\end{figure}

  We study decays in the exclusive fully reconstructed decay channel,
  \(\Sgbstpm\to\Lb\pipm_{soft}\), \( \Lb\to\Lc\pim_{b} \),
  \(\Lc\to\pKpi\)~\footnote{Charge-conjugate combinations are always
  implied unless otherwise stated.}.
  To reconstruct \( \Lb \) signal first the \(\Lc\to\pKpi\)
  candidate is built up of three tracks with \(\pt>0.4\gevc \) and with
  a fitted common vertex.  Then, the \( \Lc \) candidate within 
  \( \pm3\cdot\sigma_{M} \) range around PDG~\cite{Nakamura:2010zzi} value
  of \(M(\Lc) \) is combined with a pion \( \pim_{b} \)- track of
  \(\pt>1.5\gevc \) and the \( \Lb\to\Lc\pim_{b} \) candidate is
  subjected to a single pion \( \pim_{b} \)-track vertex fit with the
  contributing \Lc candidate constrained to its
  PDG~\cite{Nakamura:2010zzi} mass.
  Figure~\ref{fig:lb_signal} with further details provided in its caption 
  shows the reconstructed base \(\Lb\)
  signal.  This is currently the largest sample of \( \Lb \) decays,
  with \( \sim16300 \) candidates in the signal and with the ratio 
  \( S/B\approx1.8 \).
%
%
%
\par
\begin{figure}[hbtp]
%
  \includegraphics[width=0.5\textwidth]{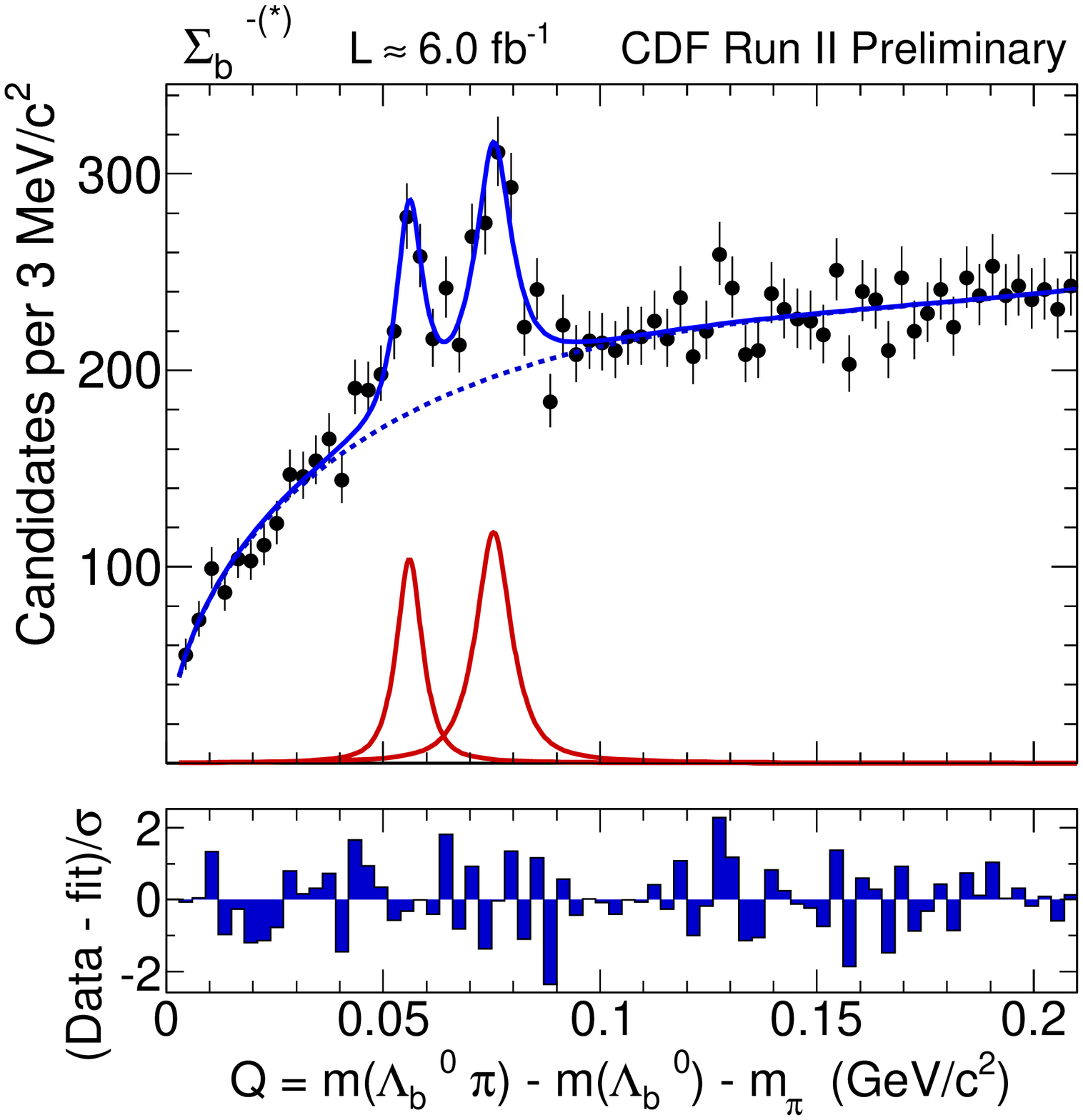}
  \includegraphics[width=0.5\textwidth]{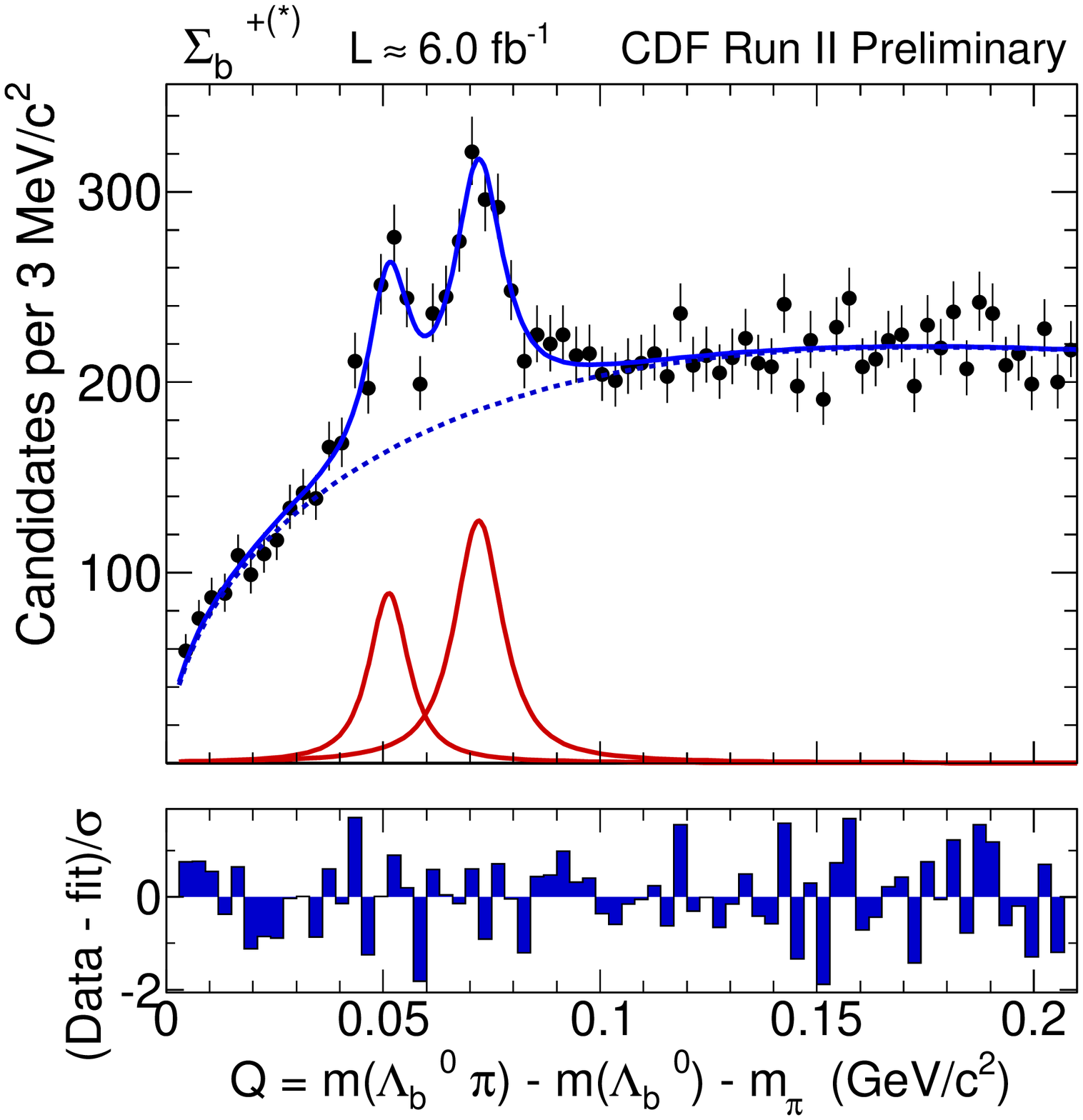}
  \caption{ The \( Q \)-value spectra, where 
            \( {Q = M(\Lb\pipm) -M(\Lb) -m_{\pipm}} \),
            are shown for \Sgbstm (left) and for \Sgbstp (right) 
            candidates with the projection of the corresponding
            unbinned likelihood fit superimposed. }
\label{fig:sgb_signals}
\end{figure}

  Finally, the \Lb candidates within \( \pm3\cdot\sigma_{M} \) window
  around \Lb signal peak are combined with a soft pion track of
  \(\pt>0.2\gevc\) and again the \(\Sgbstpm\to\Lb\pipm_{soft}\)
  candidates are subjected to a single track vertex fit.  The analysis
  is performed with the \(Q\)-value distributions, 
  \(Q = m(\Lb\pipm_{soft}) - m(\Lb) - m(\pipm)_{\rm PDG}\) where the \Lb
  candidate's resolution and most of the systematic uncertainties are
  canceled.  The signals of \( \Sgbstm \) and \( \Sgbstp \) are
  reconstructed as two peaks in the corresponding \(Q\)- value spectra
  and subjected independently to the unbinned likelihood fits.  The
  signal shapes are modeled with a non-relativistic Breit-Wigner
  function with the width modified by a \(P\)-wave factor
  \( {({\frac{p^{*}_{\pi} }{ p^{*0}_{\pi} })}^{3} } \), where
  \( p^{*}_{\pi}\) is the \( \pi_{soft}\) momentum in the \( \Sgbst \) rest frame, 
  and \( p^{*0}_{\pi} \) is the \( \pi_{soft}\) momentum in the same rest frame but 
  calculated at the resonance pole mass~\cite{jdjackson-cleo}.
  The Breit-Wigner function is further convoluted with the detector resolution described
  by two Gaussians. The background under the signals is described 
  by a formula, specifically
  \( \sqrt{(Q+m_{\pi})^{2}\,-\,thr^{2} }\cdot\,\left(C\,+b_{1}\cdot\,Q +b_{2}\cdot\,(2\cdot Q^{2}\,-1)\right)\), 
  where \( thr \) is fixed to a pion mass \( m_{\pi}=0.1396\gevcc \).
  Figure~\ref{fig:sgb_signals} shows the \( \Sgbstpm \) \(Q\)-value
  distributions with the projections of the corresponding unbinned fits
  superimposed.
\par
  The significance of the observed signals is tested against several
  null hypothesis using the log-ratio of the minima likelihoods,
  \(-2\cdot\log(\mathcal{L}_{0}/\mathcal{L}_{1})\), reached by the
  fitter for the base line hypothesis \(\mathcal{L}_{1}\) and for a
  particular null hypothesis \(\mathcal{L}_{0}\). Any signal combination 
  is by \(\gsim10.0\sigma \) significant than the sole background model.
  The most critical null hypothesis is a broad single peak fluctuating
  to the default two peak model: both \Sgbstm and \Sgbstp observed
  two-peak signal structures are individually preferred by the base line
  hypothesis at \(\gsim7.0\sigma \) level relative to such null
  hypothesis.
\par
  The measurement results are listed in a Table~\ref{tab:results}. The
  systematic uncertainties come from the residual uncertainty in the mass
  difference \(Q \)-value due to a CDF tracking momentum scale, small
  bias caused by the fitter yielding the natural width results, the
  uncertainties due to the signal and background models implemented in
  the fitter. To study the systematic effects related to the detector resolution 
  of our signals we used a large available data sample of \(\Dstarp\to\Dz\pip \).
  The momentum scale dominates the systematics on the
  fitted pole \(Q \)-values while the resolution uncertainty determines
  the systematics of the results on the widths.
\begin{table}[htb]
\begin{center}
\begin{tabular}{l|cccc}
\hline
\hline
State        & \(Q\)-value, & Absolute Mass,  & Natural Width,      & Yield, \\
             & \mevcc       & m, \mevcc       & \(\Gamma\), \mevcc  & num. of cands. \\
\hline
\hline
{\Sigbp} & {\(52.0_{-0.8-0.4}^{+0.9+0.09} \)} & {\(5811.2_{-0.8}^{+0.9}\pm1.7 \)} & {\(9.2_{-2.9-1.1}^{+3.8+1.0} \)} & {\(468_{-95-15}^{+110+18} \)} \\
{\Sigbm} & {\(56.2_{-0.5-0.4}^{+0.6+0.07} \)} & {\(5815.5_{-0.5}^{+0.6}\pm1.7 \)} & {\(4.3_{-2.1-1.1}^{+3.1+1.0} \)} & {\(333_{-73}^{+93}\pm35 \)} \\
{\Sigbstp} & {\(72.7\pm0.7_{-0.6}^{+0.12} \)} & {\(5832.0\pm0.7\pm1.8 \)} & {\(10.4_{-2.2-1.2}^{+2.7+0.8} \)} & {\(782_{-103-27}^{+114+25} \)} \\
{\Sigbstm} & {\(75.7\pm0.6_{-0.6}^{+0.08} \)} & {\(5835.0\pm0.6\pm1.8 \)} & {\(6.4_{-1.8-1.1}^{+2.2+0.7} \)} & {\(522_{-76}^{+85}\pm29 \)} \\
\hline
\hline
  & \multicolumn{3}{c}{Isospin Mass Splitting, \mevcc}  \\
\hline
 {m(\Sigbp) - m(\Sigbm)}  & \multicolumn{4}{c}{\( -4.2_{-0.9-0.09}^{+1.1+0.07} \)}\\
 {m(\Sigbstp) - m(\Sigbstm)} & \multicolumn{4}{c}{\(-3.0\pm0.9_{-0.13}^{+0.12} \) }\\
\hline
\hline
\end{tabular}
\caption{ Summary of the final results.  In all the quoted values the
         first uncertainty is a statistic one and the second one is a
         systematic.  To extract the absolute masses the best CDF mass
         measurement for \Lb~\cite{Acosta:2005mq} has been used.
}
\label{tab:results}
\end{center}
\end{table}
%
%
%
%
\par
  In a conclusion, we have measured the \Sgbstpm bottom baryons using a
  sample of \( \sim16300 \) \( \Lb \) candidates identified in 
  \( \Lb\to\Lc\pim \) mode with the \( 6\invfb \) of data collected with
  the \cdf2 detector.  More details on the presented analysis can be
  found in~\cite{sgbmeas:pubnote}.
%
%
  The first observation of \Sgbstpm bottom baryons made by CDF
  Collaboration~\cite{:2007rw} has been confirmed with the every
  individual signal reconstructed at a significance of \( \gsim7\sigma\) 
  in Gaussian terms.
%
%
  The direct mass difference measurements have been found
  with the statistical precision by a factor of \( \gsim2.3 \)
  better w.r.t. to the published results~\cite{:2007rw}
  and according to the amount of the statistics available.
  The measurements are in a good agreement with the previously found 
  results~\cite{:2007rw}. 
%
%
  The isospin mass splittings within \( I=1 \) triplets \( \Sigb \) and
  \( \Sigbst \) have been extracted for the first time. The precision of
  the experimental isospin splittings is as good as the ones quoted by
  PDG~\cite{Nakamura:2010zzi} for \( \Sigc \) states.  The \( \Sgbstm \)
  states have a higher mass value than their \( \Sgbstp \) partners
  following a standard pattern for every known isospin
  multiplet~\cite{Guo:2008ns} even though excluding their charm
  partners~\cite{Nakamura:2010zzi}, \( \Sigc \) where the supposedly
  natural order of masses within isotriplets is still
  violated~\cite{felix-wick:ichep10}.
%
%
  The natural widths of both \( \Sigbpm \) and \( \Sigbstpm \) states have 
  been measured for the first time. Given the statistical errors
  the measurements are in the agreement 
  with the theoretical expectations~\cite{widths}.
\begin{acknowledgments}
  The authors are grateful to their colleagues from the CDF {\it
  B}-Physics Working Group for useful suggestions and comments made
  during preparation of this talk.  The authors thank
  Juan~Pablo~Fernandez (CIEMAT, Spain) and Sally~C.~Seidel (Univ. of New
  Mexico, USA) for a successful cooperation and a support of this work.
\end{acknowledgments} 
\end{document}